\documentstyle[12pt,psfig,a4]{article}
\begin{document}
\thispagestyle{empty}
\noindent\hbox to\hsize{September 1996 \hfill  ULG-PNT-96-2-JRC}\\
\vskip 1.6in
\begin{center}
{\bf ELASTIC VECTOR-MESON PRODUCTION AT HERA
}\\
\vskip .3in
J.R. Cudell\footnote{cudell@gw.unipc.ulg.ac.be}
and I. Royen\footnote{royen@gw.unipc.ulg.ac.be}\\{\small
Inst. de Physique, U. de Li\`ege,
B\^at. B-5, Sart Tilman, B4000 Li\`ege, Belgium}\\
\vskip 1in
{\bf Abstract}
\end{center}
\begin{quote}
We show that the lowest-order QCD calculation in a simple model of elastic
vector-meson production does reproduce correctly the ratios of cross sections
for $\rho$, $\phi$ and $J/\psi$, both in photoproduction and in high-$Q^2$
quasi-elastic scattering.  The dependence of the slopes on the mass of
the vector meson is reproduced as well. We examine the lower-energy data,
and find that the energy dependence of the cross section does not depend on
$Q^2$, but may depend on $m_V$.
\end{quote}
\newpage
\newcommand{\w}{\hat w^2}
\newcommand{\mv}{$M_V$}
Elastic vector-meson production opens a precious window on the interface
between
perturbative QCD (pQCD) and non-perturbative hadronic physics,
and is complementary to
deep-inelastic scattering (DIS).  Indeed, DIS
has been until recently the triumph of perturbative ideas, whereas
elastic processes were mostly treated through non-perturbative methods.
These two processes now meet at HERA, where pQCD and Regge models
have to be merged to obtain a full understanding of the data. Elastic
vector-meson production has the extra advantage of containing  by
definition two scales, the mass of the produced vector meson, $M_V$, and
the  off-shellness of the photon, $Q^2=-q^2$. We shall see that both
dependences
can be understood through
a lowest-order calculation.

The data for $\gamma^* p\rightarrow V p$, which both H1 \cite{dataH,dataHp} and
ZEUS \cite{dataZeus}
have obtained, for $V=\rho$,
$\phi$ and $J/\psi$, exhibit the following main features:
(i) the $Q^2$-distribution of the cross section is shallower for
$J/\psi$  than for $\rho$, the two cross sections becoming comparable
around $Q^2=10$ GeV$^2$ and (ii)
the $t$-slopes of the differential elastic cross sections
depend on the mass of the
produced meson, and become shallower as the mass increases.

Several models have been proposed to describe this process. Originally,
\break Donnachie and Landshoff (DL) \cite{DL1} extended their soft-pomeron
model to
predict $\rho$ production at EMC \cite{datalowe}. This model works well there,
but it has not been applied to predict the ratios of produced vector
mesons. In photoproduction, DL preferred to resort
to the quark counting rule to predict the ratios of cross sections~\cite{DL2},
and reached the conclusion that it does not work perfectly, but argued that
the violations were reasonable.
The transition to perturbative QCD was first introduced by DL \cite{DL3}, who
noticed the analogy between the pomeron
expressions and two-gluon exchange, in the transverse case. They used a
``constituent gluon'' propagator and two-gluon exchange to model the pomeron.
This analogy was pursued by one of us \cite{cudell}, who showed that such
a model can give reasonable agreement with EMC data. This was later confirmed
by NMC \cite{datalowe}. The final step to pQCD was performed by Ryskin
\cite{Ryskin}
who observed that at high-$Q^2$ and high \mv, the effective intercept
should be analogous to that found in $x g(x)$. This was later confirmed by
Brodsky et al. \cite{Brodsky}. So far, there has not been a model applied to
the
full range of masses and $Q^2$: this is the object of this letter.

A model for exclusive vector-meson production must include three sub-mo\-dels,
for which we shall adopt the simplest ones:
the transition $\gamma^*\rightarrow V$ is described by a zero Fermi momentum
wavefunction, the colour-singlet
exchange is modeled \`a la Low-Nussinov \cite{Low}, and we shall only
consider the constituent quarks of the proton.
There are in principle 72 diagrams contributing to the amplitude:
the gluons can be hooked 4 different ways to the quarks of the
vector meson, 9 different ways to the quarks of the proton, and both the
direct and the $s\leftrightarrow u$ channels contribute to the amplitude
in the high-$s$ limit. As we shall explain however, the calculation of each
part of the amplitude can be greatly simplified, so that one needs to
calculate only the two diagrams shown in Fig.~1.
\begin{figure}
\centerline{
\hbox{
\psfig{figure=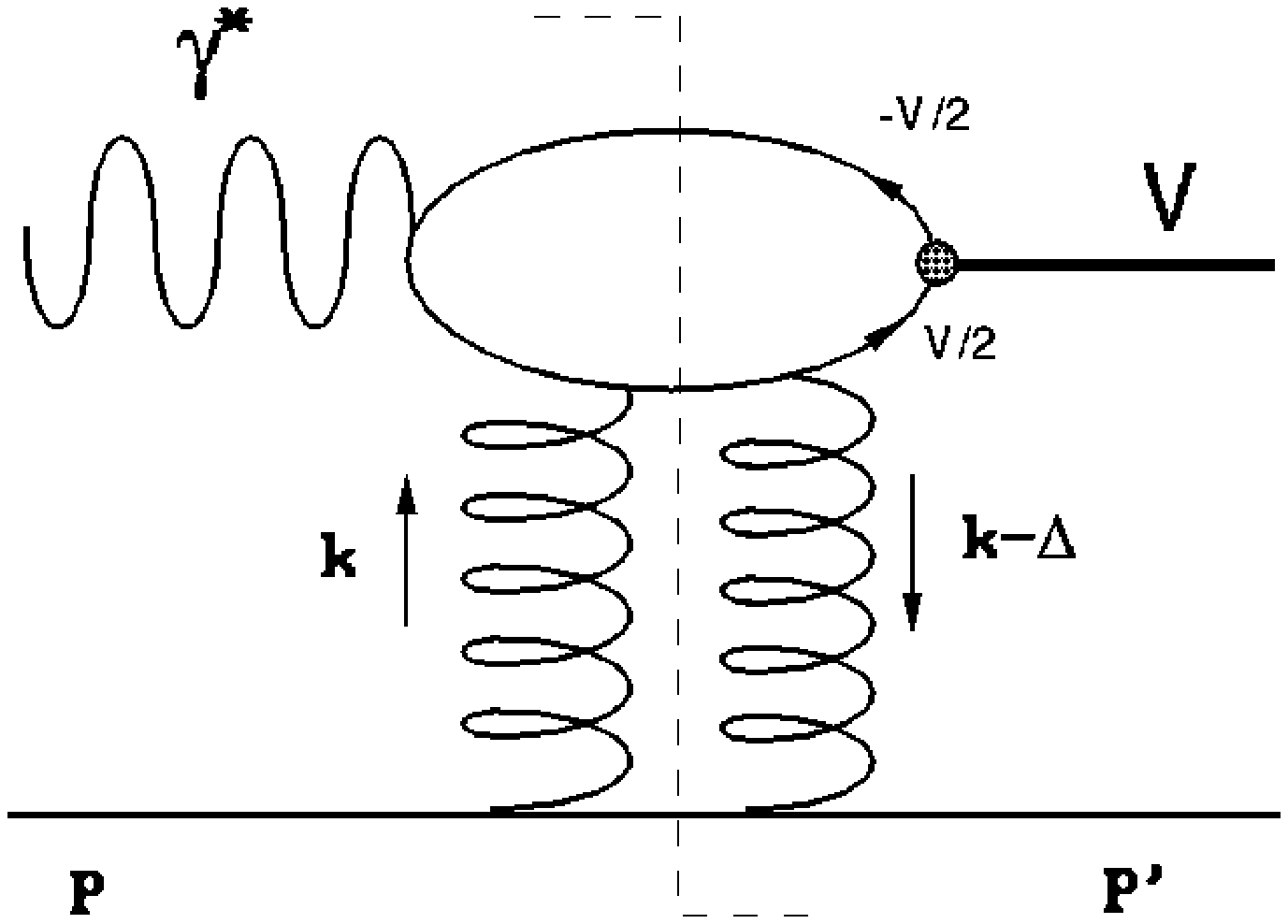,bbllx=2.5cm,bblly=13cm,bburx=17cm,bbury=23cm,width=5cm}
\hglue 1cm
\psfig{figure=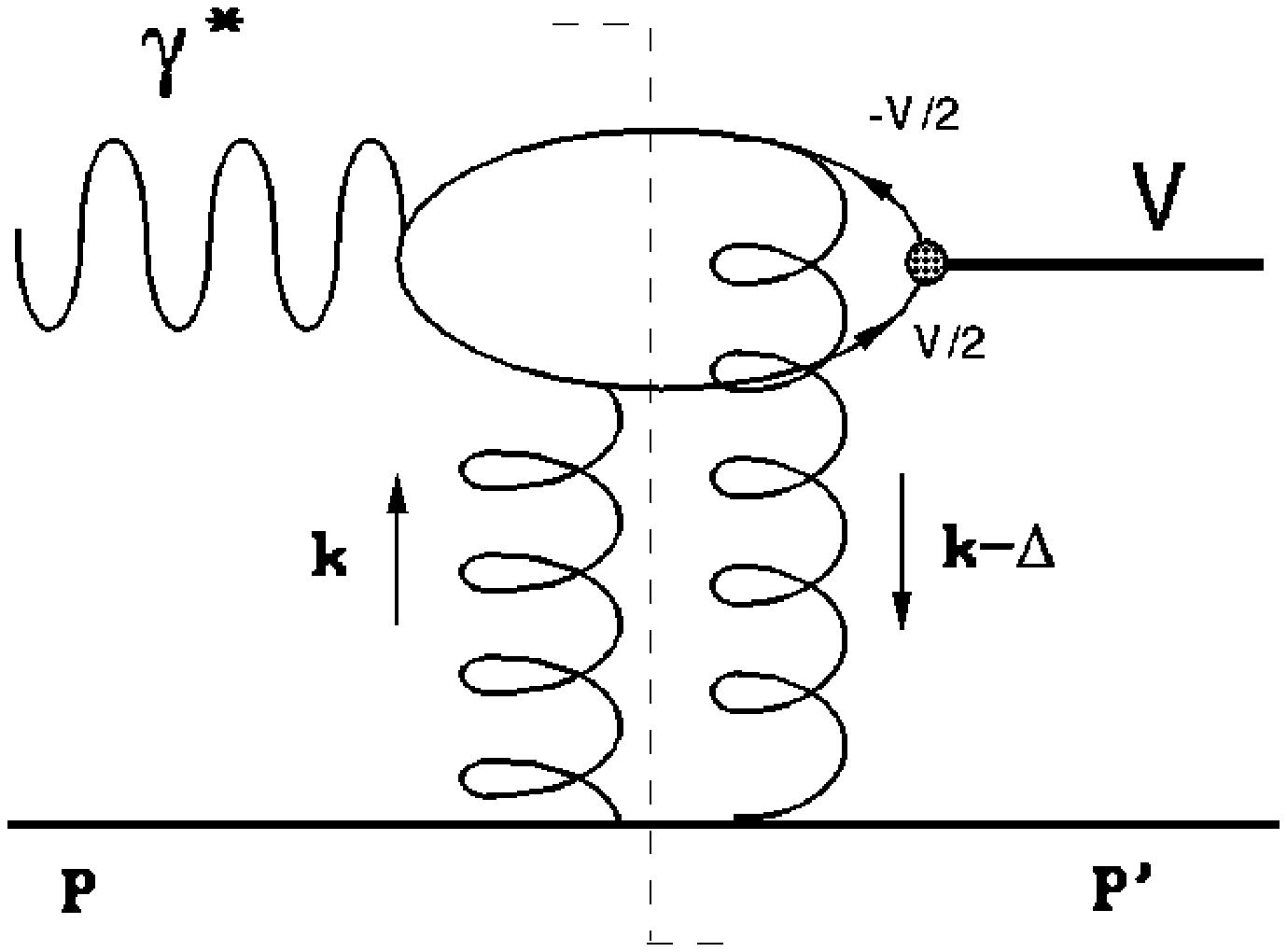,bbllx=2.5cm,bblly=13cm,bburx=17cm,bbury=23cm,width=5cm}}}
\begin{quote}
{\small Figure~1:  The two diagrams accounting for the transition
$\gamma^*\rightarrow V$.
The dashed line represents the cut which puts the intermediate state on-shell.
}\end{quote}
\end{figure}

Let us first consider the kinematics of the process, which is
spelled out in Fig.~1.
Let $P$ and $q$ be respectively the 4-momenta of the proton and the photon, and
$P-\Delta$ and $q+\Delta$ the momenta of the final-state proton and
vector meson. We work in the high-$w^2$ limit, with $w^2=(P+q)^2$.
The on-shell condition for the proton and the vector meson
imply that $\Delta$ is transverse to order $1/w^2$: $\Delta\approx\Delta_T$
with $\Delta_T\cdot P=\Delta_T\cdot q=0$, and
$|\Delta_T^2|\approx|\Delta^2|=-t$.

As it will turn out, the
imaginary part of the amplitude is proportional to $w^2$. Crossing
symmetry and analyticity then imply that the amplitude is purely imaginary,
up to terms of order $1/w^2$, because the exchange is $C=+1$. Hence in the
following, we shall calculate only the imaginary part, using
Cutkovsky's rules, and putting intermediate states on-shell, as shown in
Fig.~1.
The quarks that
make the vector meson are in the direction of $q$, whereas those that make
the proton are parallel to $P$. The intermediate states come from
the absorption of one gluon. In order for these states to be on-shell, the
gluon must have vanishing components both in the direction of $P$ and $q$,
hence the gluon momenta are transverse to order $1/w^2$:
$k\approx k_T$, with $k_T.P=k_T.q=0$.

Hence the gluon momentum is essentially transverse, and furthermore
we do not need to worry about the (purely real) crossed diagrams.
The transversality of the gluons implies that the embedding of a process
at the quark level into a proton is particularly simple.
We follow here Gunion and Soper \cite{GS}, and choose to represent the
proton by a constituent model. This naturally leads to the quark counting
rule, and has the extra advantage that the form factor of the proton is to
a large extent measured.
Indeed, when both gluons hit the
same quark line, the form factor is given by the Dirac elastic
form factor, measured in $ep$ scattering:
\begin{equation}
{\cal E}_1(t=\Delta^2)\approx \frac{(3.53-2.79t)}{(3.53-t)(1-t/0.71)^2} \\
\end{equation}
If the gluons hit different quark lines, then the form factor is not known,
but we know that infinite wavelength gluons must average out the colour
of the proton, hence we know that this form factor has to cancel the
infrared singularity that would result from the pole in the gluon propagator,
and therefore it must
reduce to ${\cal E}_1(t)$ when either gluon goes on-shell.
We choose to parametrise this form factor as:
\begin{equation}
{\cal E}_2(k,k-\Delta)={\cal E}_1(k^2+(k-\Delta)^2+c\ k\cdot (k-\Delta))
\end{equation}
There are theoretical arguments \cite{CN} as well as phenomenological
ones \cite{CH} which lead to the conclusion that $c\approx -1$.

The rule is then to calculate the process at the quark level. This leads to
an integral over the transverse momenta of the gluons. One then introduces
the difference of form factors $3 ({\cal E}_1-{\cal E}_2)$ into the integral
to get the same process at the proton level, thereby reducing the number of
diagrams by a factor 9. For the lower trace, and in the high-$w^2$ limit,
we need to keep only the leading terms in $p$, the momentum of the quark,
hence the trace along the quark line in the square of the
amplitude, is given by $(4 p^\alpha p^\beta) \times (4 p^{\alpha'} p^{\beta'})$
(including a factor of 1/2 for spin averaging),
with $p$ the momentum of the quark inside the proton, and $\alpha^{(_{'})}$ and
$\beta^{(_{'})}$ the indices of the gluons.
Thus we can treat the process at the level of the amplitude, without
the need to square it, provided that we write the contribution of
the lower quark line as
$4 p^\alpha p^\beta$.

For the vector meson, we use a different model than for the proton, as we
want to take into account the mass of the meson.
The vector meson is modeled \cite{Horgan} by
its lowest Fock state, with no Fermi momentum, which implies  $m_q=m_V/2$, and
the $V\bar q q$ vertex, including the two quark propagators, is given by:
\begin{equation}
\Phi_V=C' \ (m_q-\gamma.v)\ \gamma.e\ (m_q+\gamma.v)
= C\ \gamma.e\ (m_V+\gamma.V)
\end{equation}
where $v$=${V\over 2}$ is the quark momentum within the vector meson
and  $C\equiv\sqrt{f_V/24}$ is the normalisation that
reproduces the vector
meson decay rate, with\break \hbox{$f_V\approx 0.025 ({\rm GeV})\ m_V$} the
vector-meson decay
constant squared.
This effective vertex includes the propagators of the quark lines
flowing into it. Reversing the direction of the quark current gives the
same contribution, therefore we end up with only 2 diagrams to calculate -
those
shown in Fig.~1.

The traces corresponding to the upper bubble, dotted into
$p_\alpha p_\beta$, are:
\begin{eqnarray}
\small {\cal T}_1&=&Tr[\rlap{\slash}e\ (\gamma.V+m)\ \rlap{\slash}{p}\
(\gamma.(q-{v
}+k)+{m_q})\ \rlap{\slash}p\ (\gamma.(q-{v
})+{m_q})\ \rlap{\slash}\epsilon]~~~~~~\\
{\cal T}_2&=&Tr[\rlap{\slash}e\ (\gamma.V+m)\ \rlap{\slash}p\
(\gamma.(q-{v
}-k+\Delta)+{m_q})\ \rlap{\slash}\epsilon\nonumber\\
&&\times (\gamma.(-{v
}-k+\Delta)+{m_q})\ \rlap{\slash}p]\end{eqnarray}
with $e$ and $\epsilon$ the polarisations of the vector meson and of the
photon.
The propagators of the off-shell quarks are:
$ p_1=-(Q^2+m_V^2-t)/2$ and $ p_2= ( m_V^2 + Q^2 -t-4k.q)/2$.
The answer is then proportional to:
${\cal T}={\cal T}_1/p_1+{\cal T}_2/p_2$.
This answer is explicitly gauge invariant:
\renewcommand{\thefootnote}{\dagger}
substituting $\epsilon=q$ in ${\cal T}$ gives 0\footnote{Note that this was not
the case for the expressions previously published in \cite{cudell}.
The difference at high-$Q^2$
amounts to a factor of 2 in the amplitude,
presumably the discrepancy pointed out in \cite{Laget}.}.

In the transverse case, the leading dot product is simply $\w\equiv
(p+q)^2\approx 2 p.q$.
For longitudinal polarisation, further contributions appear:
$\epsilon_L.p\approx{-\w}/({2\sqrt{Q^2}})$ and
$e_L.p\approx{-\w}/({2m_V})$.
Keeping track of these, the leading term is proportional to:
\begin{eqnarray}
{\cal T}&=& {2m_V\ k\cdot(k-\Delta)\over t - m_V^2 - Q^2 + 4\ k\cdot(k-\Delta)}
\nonumber\\
&& \times\left({2\Delta.\epsilon\ e.p  +2 q.e\ \epsilon.p - \epsilon.e
\ \w\over (m_V^2 + Q^2 - t)}-{2\over \w}
\ \epsilon.p\ e.p \right)\end{eqnarray}

Putting everything together, we obtain the following expression for the
amplitude:
\begin{equation}
{\cal A}=-i\ \alpha_S^2\ g^{elm}_V \frac{2}{3}\ \sqrt{\frac{m_V\ f_V}{24}}
\int\ \frac{d^2k_T}{k^2(k-\Delta)^2}
\ 3[{\cal E}_1(t)-{\cal E}_2(k,k-\Delta)] \times 32{\cal T}
\label{amplitude}\end{equation}
where $g^{elm}_V=\xi \sqrt{4\pi \alpha_{elm}}$ is the electromagnetic coupling
of the different vector mesons: $\xi=\frac{1}{\sqrt{2}}$ for the $\rho$,
$-1/3$ for the $\phi$ and $2/3$ for the $J/\psi$.
For the various possible helicities, Eq.~(\ref{amplitude}) gives:
\begin{eqnarray}
{\cal A}(T\rightarrow T)&=&i\w\frac{64\alpha_S^2\ g^{elm}_Vm_V\sqrt{m_V\
f_V}}{\sqrt{6}} \nonumber\\
&&\hskip -2.3cm \times \int \frac{d^2k_T}{k^2(k-\Delta)^2}
{[{\cal E}_1(t)-{\cal E}_2(k,k-\Delta)]
\ [k\cdot(k-\Delta)]\over \left(t - m_V^2 - Q^2 + 4\ k\cdot(k-\Delta) \right)\
(m_V^2 + Q^2 - t)}\\
{\cal A}(L\rightarrow L)&=&{\sqrt{Q^2}\over m_V}\times {\cal A}(T\rightarrow
T)
\end{eqnarray}
The helicity violating amplitude ${\cal A}(L\rightarrow T)$ is suppressed
by $1/\w$.
As previously advertised, this answer is proportional to $\w$, therefore to
$w^2$,
and is thus purely imaginary. The resulting cross section is independent
of $w^2$. Clearly this model cannot say anything about the energy dependence
of the cross section. We shall assume that it comes in as a factor, $R$,
and check whether the latter is mass- or $Q^2$-dependent. Note that we can
only determine the value of that factor times $\alpha_S^2$. In the following,
we shall let $\alpha_S$ run with the off-shellness of the gluons, and freeze it
at some value $\alpha_S^0$. However, as we shall see later, the dominant
contribution comes from gluons of small off-shellness, and the results we
obtain are identical to fixed-coupling results for $\alpha_S=\alpha_S^0$.

The differential cross section is given by:
\begin{equation}
{d\sigma\over dt}={d\sigma_T\over dt}+\varepsilon {d\sigma_L\over dt}
={R\over 16\pi(\w)^2} \left[|A(T\rightarrow T )|^2+\varepsilon|A(L\rightarrow L
)|^2\right]
\end{equation}
with $\varepsilon$ the polarisation of the photon beam: $\varepsilon\approx 1$
at HERA
and $\varepsilon\approx 0.75$ at NMC.

\begin{figure}
\centerline{
\psfig{figure=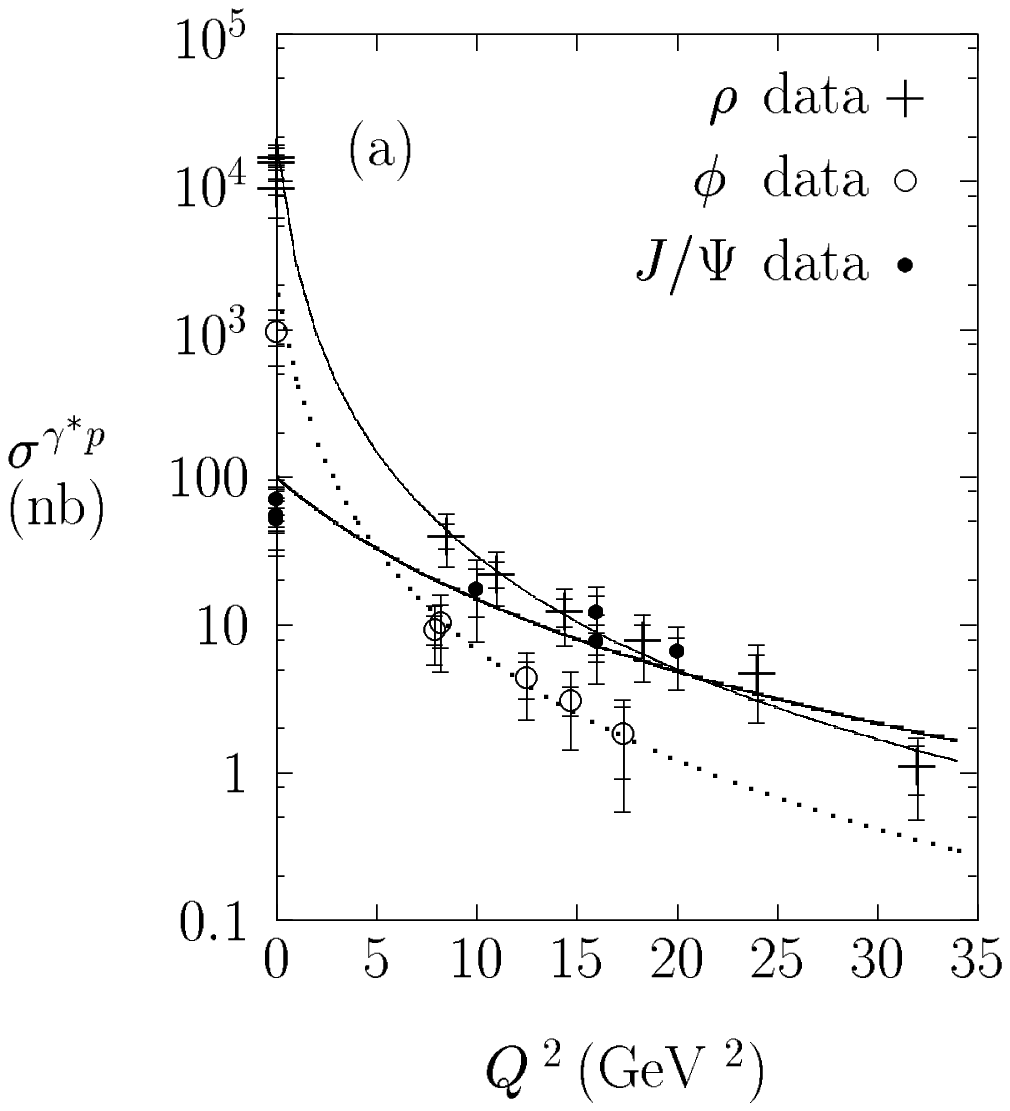,height=6cm}\ \ \ \ \
\psfig{figure=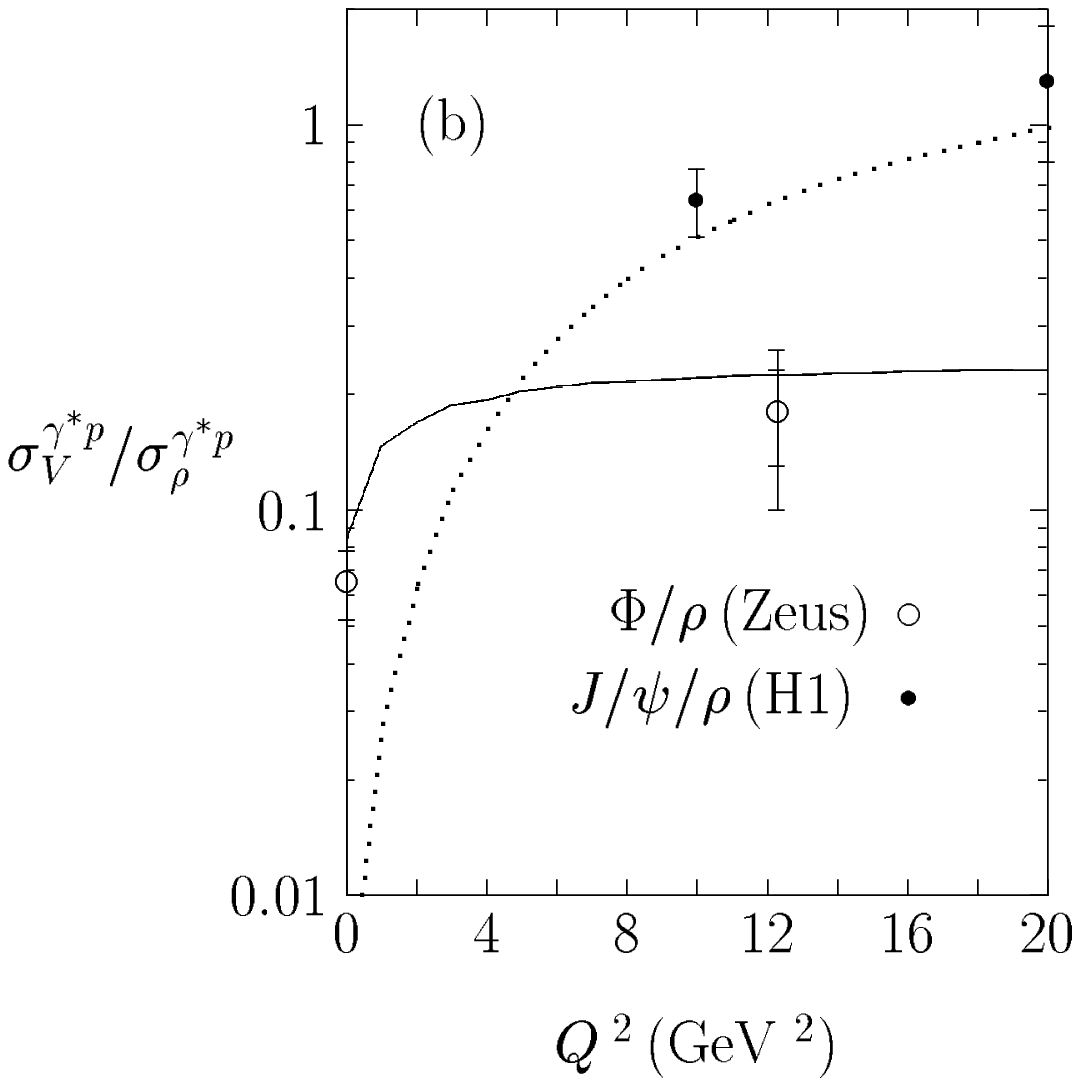,height=6cm}}

\begin{quote}{\small Figure~2: (a) Cross sections as functions of $Q^2$,
compared
with data from H1~\cite{dataH,dataHp} and ZEUS~\cite{dataZeus}
at $<w>\approx$100~GeV, (b) Ratio of cross sections as functions of $Q^2$
at \hbox{$<w>\approx$100~GeV}, compared with data from H1 \cite{dataH,dataHp}
and Zeus
\cite{dataZeus}.
}\end{quote}
\end{figure}

We first give the results that we obtain
for the various cross sections measured by
ZEUS and H1. First of all, we show in Fig.~2 the dependence on $Q^2$ and
$m_V$ of the integrated elastic cross section $\sigma(Q^2)$.
We see that a common (Regge) factor is consistent with the data
taken at HERA, as shown in Fig.~2(a). We insist on the fact that this factor
is independent both of $Q^2$ and of $m_V$, as one would expect within
Regge theory.  Selecting only high-$Q^2$ data leads to a best value
$R\alpha_S^2=0.6$, with a $\chi^2/d.o.f.=0.39$. We do not find that
$R\alpha_S^2$ varies significantly within the energy range of HERA.
Although we see no reason why our model should work in photoproduction,
it turns out that our curves do go
through the photoproduction points. Including these in the fit brings
 $R \alpha_S^2$ to 0.55.
\begin{center}\vbox{
\begin{tabular}{|c|c|c|}\hline
meson& $A$ (nb)&n \\ \hline
$\rho$&5534&2.5  \\
$\phi$&1035.9&2.4  \\
$J/\psi$&269.53&1.5\\
\hline
\end{tabular}
{}~\\
{}~\\
{\small Table 1: Result of a fit of $\sigma(Q^2)$ to $A (Q^2)^{-n}$, for
5 GeV$^2<Q^2<25$ GeV$^2$.}}
\end{center}

We give in
Table 1 the result of a fit to a power of $Q^2$ at large $Q^2$. Although
asymptotically all cross sections behave like $(Q^2)^{-3}$, in agreement with
\cite{Ryskin, Brodsky}, we see that the data collected at HERA are not yet
in that asymptotic regime. Note that our calculation holds only for $Q^2<<\w$,
and hence it is not clear whether the asymptotic regime will be reachable at
HERA. Our model is equivalent to  those of refs~\cite{Ryskin, Brodsky}, with
a gluon distribution corresponding to 2-gluon exchange $g(x)\sim 1/x$.
We see that here the $Q^2$ dependence is already reproduced, hence there
is no room for an extra dependence coming from the gluon distribution,
and the use of the asymptotic formula \cite{MRT} is misleading.

In Fig.~2(a), we show both the systematic and the statistical errors.
In the ratio of cross sections, some of the systematic uncertainties
cancel, and the reproduction of that ratio is a more stringent test of
our model, especially as the normalisation then drops out of our prediction.
We show in Fig.~2(b) the result of such a comparison.
Again, we see that our model fares well, even in photoproduction.

Hence we can understand both the $m_V$- and the $Q^2$-dependence
of the cross sections. One might object that this is because these
are concentrated at low $t$, and argue that the $t$-dependence has
to be wrong, as this is one of the well-known problems of perturbative
calculations applied to diffractive scattering.

The behaviour of the slopes as a function of $m_V$ and $Q^2$ can be understood
as follows. We can approximate the proton form factor ${\cal E}_1(t)
-{\cal E}_2(k,k-\Delta)$ as being proportional to
${\cal E}_1(t) [k\cdot (k-\Delta)]$, using the fact that $F_1$ is close to an
exponential, and expanding for small $k\cdot (k-\Delta)$. The amplitude
of Eq.~(\ref{amplitude}) can
then be written as $C(m_V,Q^2) \sqrt{R(t)} {\cal E}_1(t) {\cal F}\left({t\over
Q^2+m_V^2}
\right)$, with $C$ a constant with respect to $t$, and ${\cal F}$
a calculable function.
This means that the logarithmic derivative of $d\sigma/dt$ becomes:
\begin{equation}
b(t)= {d{\cal E}^2_1\over {\cal E}_1^2 dt}
+ {dR\over R dt}
+ {1\over (m_V^2 +Q^2)} {d{\cal F}^2\over {\cal F}^2dx}{\Big |}_{x={t\over
Q^2+m_V^2}}\label{slopes}
\end{equation}
Thus the slope is approximatively made of three terms: one corresponding to
the proton response, one to the pomeron response, and one to the response of
the loop which converts the photon into a vector meson. Only the latter
depends on $m_V$ and $Q^2$, and we see that it decreases rather fast with
both of these factors. At large $Q^2+m_V^2$, it becomes negligible, and only
the first two responses matter. This means that this kind of model predicts
that all the slopes have to reach the same asymptotic value. This value is
about
4~GeV$^{-2}$.

This variation of the slopes with $Q^2+m_V^2$ enables us to reproduce
the measurements of HERA: in photoproduction, we obtain
$1/<t_\rho>=8.75$~GeV$^{-2}$, $1/<t_\phi>=7.35$~GeV$^{-2}$ and
$1/<t_{J/\psi}>=4.55$~GeV$^{-2}$, whereas the $\rho$ slope is already down
to 5~GeV$^{-2}$ for $Q^2=5$~GeV$^2$. This is inconsistent with the data from
H1, but agrees with those from Zeus.
We want to point out that the experimental evaluation of the
slope demands that the cross sections be exponential in $t$.
We have quoted our results for $1/<t>$, which in the case of an
exponential fall-off $e^{bt}$ is equal to the logarithmic slope $b$.
The fact that the differential cross section is not an exponential makes
the comparison with data difficult.
To illustrate the effect of the curvature, we compare in Fig.~3 our results
with the data for $\rho$ photoproduction in H1. We see
that although the slope is supposed to be 10.9, our curve reproduces the data
fairly well, with a $1/<t>$ of 8.75 GeV$^{-2}$. Hence the different definition
of the slope leads to a possible systematic correction of
about 2 GeV$^{-2}$.
No matter which
model is used, $d\sigma/dt$ is {\it not} an exponential, and we urge the
experimentalists to quote a $<t>$ instead of a logarithmic slope.
In fact, our model works although it assumes no shrinkage. There is room in the
$\rho$ case to add a slope coming from the Regge factor $R$, which would
contribute
about 4.6 GeV$^{-2}$ to $b$ for a pomeron slope $\alpha'=0.25$ GeV$^{-2}$. One
would
then expect a similar contribution in the $J/\psi$ case. Hence if the H1
measurements of $\rho$ slopes are correct, it is likely that the $J/\psi$
numbers have a background problem: inelastic contributions would significantly
reduce the slope.

The only problem at HERA seems to be the helicity structure of the
cross section. The data support the prediction that helicity is conserved,
but the ratio $\sigma_L/\sigma_T$ does not follow the results of our model.
This ratio has to behave as $Q^2$ for near-shell photons,
as a consequence of gauge invariance. Our model fulfills this requirement,
but it also predicts that this linear behaviour continues for all $Q^2$. The
data, on the other hand, seem to indicate that the ratio reaches a plateau
around 2 at high $Q^2$. This would indicate that our high-$Q^2$ transverse
cross section is wrong. This is indeed possible:
we have assumed that the meson wave function is dominated at high $Q^2$ by
configurations in which both quarks have equal momenta. In fact, it
is likely that further configurations exist \cite{Zak} which would give
additional contributions to the transverse cross section.
This may account
for the fact that this model does not reproduce the ration $\sigma_L/\sigma_T$
measured at HERA.
We plan to examine the role of Fermi momentum in a later paper.
Another interesting possibility has recently been pointed out \cite{MRT}:
the form factor describing the recombination of quarks into a $\rho$ meson
is taken to be 1 once the quarks are restricted to the mass region of
the $\rho$, and it is argued that the process will remain largely elastic.
The ratio $\sigma_T/\sigma_L$ is then shown, in the leading-log approximation,
to be distorted, at $t=0$,
by the $x$ and $Q^2$ dependence of the gluon distribution, in such a manner
that it reproduces the data.
This is the only model so far which manages to reproduce
the observed ratio, but it can be used only for $\rho$ mesons
at $t=0$, and heavily relies on
the perturbative evolution of the gluon distribution.
\begin{figure}
\centerline{
\psfig{figure=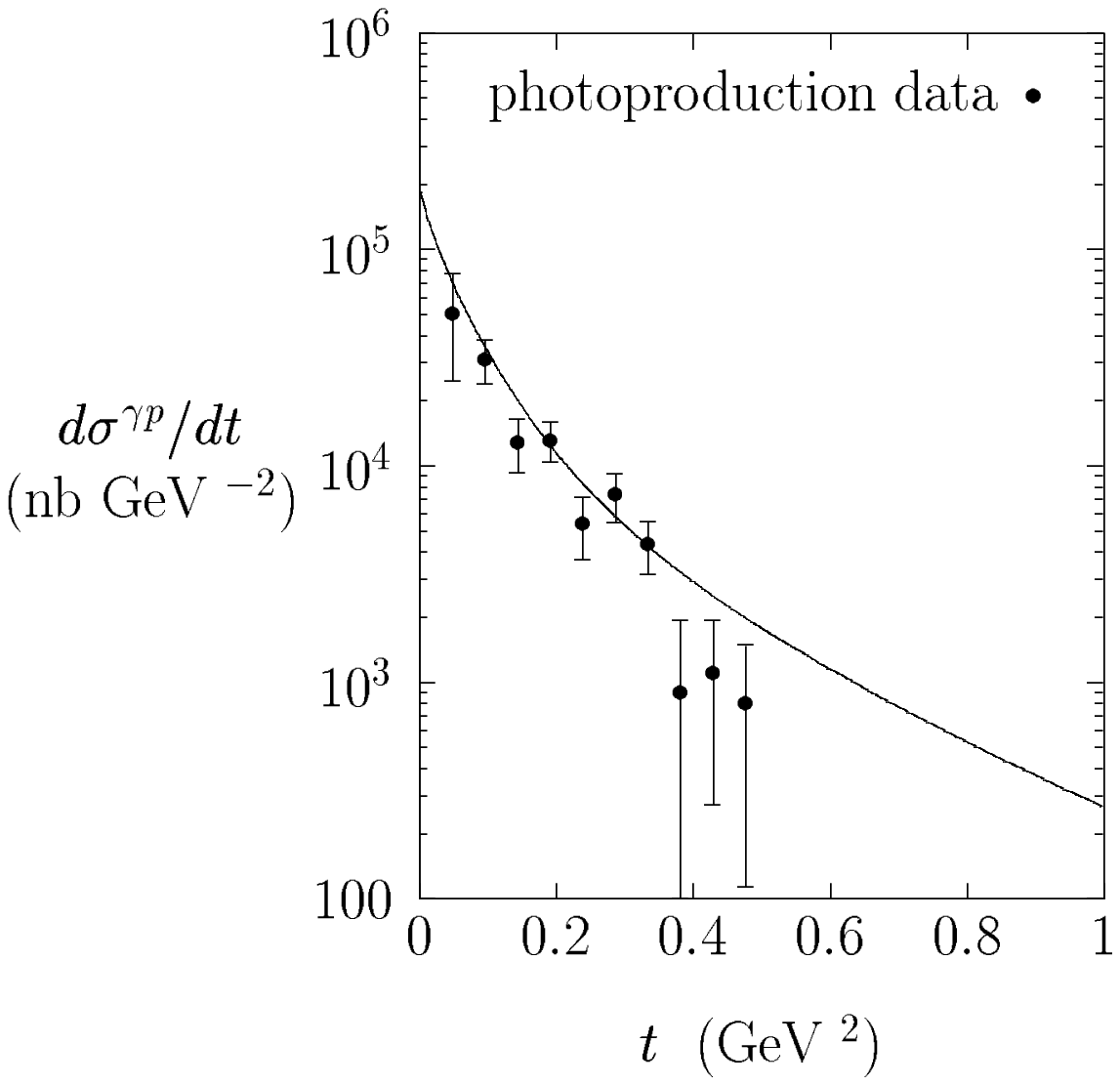,height=6cm}}
\begin{quote} {\small Figure~3:  Photoproduction differential cross section
$d\sigma/dt$, compared with H1 data \cite{dataH}.
}\end{quote}
\end{figure}

Finally, we can now examine the $w^2$-dependence
of the cross sections. We have seen that at HERA, the Regge factor does not
seem to
depend either on the meson mass or on $Q^2$. This is clearly reminiscent
of the behaviour expected from a simple pole.
We adopt the same philosophy when fitting to lower-energy cross sections from
EMC and NMC \cite{datalowe}, and write $R=s^{2\alpha_0-2}$, $\alpha_0$ being
the intercept of the pomeron.
We show in Fig.~4 the curves which correspond to $\alpha_0= 1.16$. This
value of the soft intercept is rather high, but may not be entirely excluded by
fits to total cross
sections, especially once the effect of unitarisation is taken into account
\cite{CK,CMG}. We see that the fit is reasonable for the $\rho$ and the $\phi$,
but fails for the $J/\psi$. The fact that the $\rho$
data is somewhat high can be understood
from the fact that there are contributions from
lower trajectories to the $\rho$ production cross section. The interference
between $a/f$ exchange and pomeron exchange could contribute as much as 20\%,
hence a high $\rho$ cross section. In the $J/\psi$ case, one has to realise
that the calculation we have presented here is valid if $Q^2+m_V^2<<\w$. This
is not the case for the $J/\psi$ at EMC.
Besides, the data for $J/\psi$ production from EMC is not subtracted for any
diffractive background, and can only constitute an upper bound
on the elastic cross section. It is likely that there is a contamination
of the order of a factor 2 from inelastic contributions, as the inelastic
background is larger than in the $\rho$ case, for which the EMC data were
severely contaminated \cite{datalowe}.
\begin{figure}
\centerline{
\psfig{figure=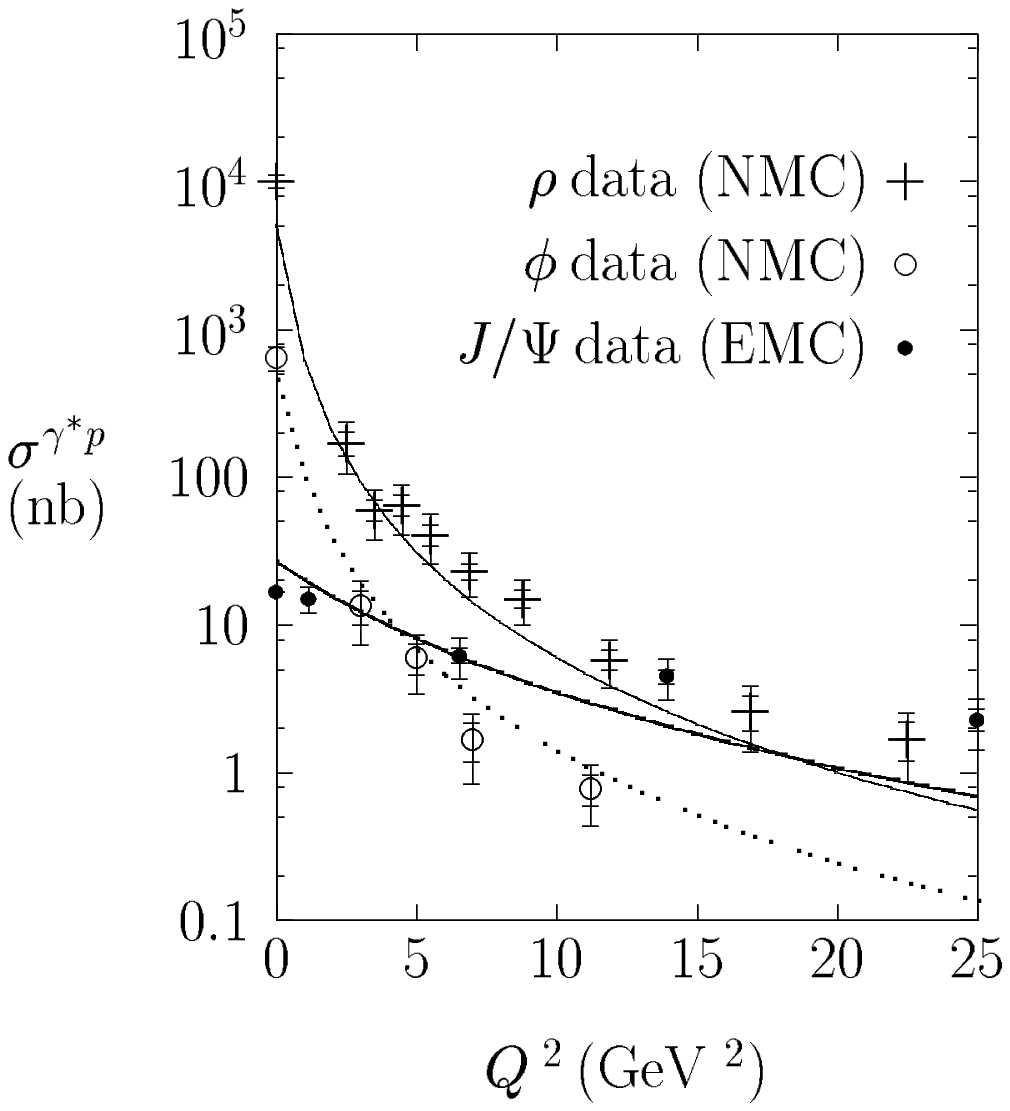,height=6cm}}
\begin{quote} {\small Figure~4:  Cross sections compared to lower-energy EMC
and NMC data~\cite{datalowe}, at \hbox{$<w^2>\approx$200~GeV$^2$}.
}\end{quote}
\end{figure}

To check whether the behaviour is indeed that of a simple pole, we can fit the
factor $R$ separately for each meson. We give in Table 2 the best factors, and
the corresponding intercepts, given the value of $R\alpha_S^2$ at HERA. We do
not give the results for the $J/\psi$ as
it is impossible to obtain a good fit to the data. One sees
that the $\phi$ and $\rho$ data require somewhat different factors,
This can be seen more clearly when one notices that
the ratio $r$ of cross sections has changed when going from HERA ($r=0.18$)
to NMC ($r=0.12$). Hence there may be a problem with the smallness of the
$\phi$ cross section at NMC, which seems to imply an intercept too large to be
compatible with that of a soft pomeron.
On the other hand, there is no sign in either $\rho$ or $\phi$ data of a $Q^2$
dependence of the intercept.

\vbox{
\begin{center}
{}~\\~\\ \begin{tabular}{|c|c|c|c|c|}\hline
meson&$R\alpha_S^2$&$\chi^2/d.o.f.$&energy (GeV)&intercept \\ \hline
$\rho$&0.4&0.6&$40<\nu<180$&1.10\\
$\phi$&0.268&0.138&$40<\nu<180$&1.206 \\
combined&0.32&1.19&$40<\nu<180$&1.161\\
\hline
\end{tabular}
{}~\\
{}~\\{\small Table 2: best values of $R\alpha_S^2$ at NMC.}
\end{center}}

Before concluding,
we must mention that although the above looks like a
successful perturbative calculation, most of the contribution to
the total cross section comes from the infrared region.
Imposing a cut-off $|k^2|, |(k-\Delta)^2| > 1$~GeV$^2$ reduces the cross
section by a factor 10.
This dominance of the infrared region justifies a posteriori our
choice of scale in $\alpha_S$: we see no theoretical reason to make
it run with either $Q^2$ or $m_V^2$, as these scales are unrelated to
the off-shellnesses of the gluons entering the vertices.
The dominance of the infrared region confirms the results
of \cite{CDL}, where it is argued that the BFKL perturbative resummation
is dominated by the non-perturbative region.

The simplest modification to the infrared region follows the ideas
of \break Landshoff and Nachtmann
\cite{LN}, which have recently been further motivated by lattice studies
\cite{UKQCD}, that the gluon propagator needs to be modified
at low $k^2$,  taming its
behaviour to something softer than a pole. One of the
main effects of that modification is that the differential cross section
becomes much more linear, and that as a consequence $<t>$ becomes bigger.
We have checked that such a model has all the other features of
the one detailed above. The only major difference is that the slopes reach
their asymptotic values much sooner, but it remain impossible to
accommodate both the
$\rho$ shrinkage and the $J/\psi$ data.

To sum up, we have shown that many features of elastic vector-meson production
at HERA can be understood in a simple QCD model. We have also seen
that comparison with lower-energy data proves somewhat problematic.
The HERA data seem to indicate that the Regge factor does not depend either
on $Q^2$ or $m_V$. The comparison with NMC $\rho$ data points to
a soft pomeron intercept, whereas the $\phi$ data
would favor a larger one. We do not believe that much can be concluded from the
$J/\psi$ data, which seem to have an inelastic background. Following \cite{CH}
we plan to model this background in a further publication.
\vfill\break
{\noindent \Large \bf Acknowledgments \hfil}\\
\indent We thank the H1 collaboration for making their data available
to us, and in particular Pierre Marage, R. Roosen, P. Van Esch and
B. Clerbaux for their hospitality and for numerous
fruitful discussions. We also thank Peter Landshoff, Joseph Cugnon
and  Boris Kopeliovich for their comments and suggestions.

\end{document}